\documentclass[twocolumn,preprintnumbers,amsmath,amssymb,superscriptaddress]{revtex4}
\usepackage{graphicx}
\usepackage{dcolumn}
\usepackage{bm}
\usepackage{soul}
\usepackage{color}
\usepackage{epstopdf}
\usepackage[version=3]{mhchem}
\usepackage{lipsum}
\usepackage[outercaption]{sidecap}
\usepackage{floatrow}
\usepackage{hyperref}

\begin{document}


\title{Deep Learning for The Inverse Design of Mid-infrared Graphene Plasmons}

\author{Anh D. Phan}
\email{anh.phanduc@phenikaa-uni.edu.vn}
\affiliation{Phenikaa Institute for Advanced Study, Faculty of Information Technology, Materials Science and Engineering, Phenikaa University, Hanoi 12116, Vietnam}
\affiliation{Artificial Intelligence Laboratory, Phenikaa University, Hanoi 12116, Vietnam}
\affiliation{Department of Nanotechnology for Sustainable Energy, School of Science and Technology, Kwansei Gakuin University, Sanda, Hyogo 669-1337, Japan}
\author{Cuong V. Nguyen}
\affiliation{Artificial Intelligence Laboratory, Phenikaa University, Hanoi 12116, Vietnam}
\author{Pham T. Linh}
\affiliation{Graduate University of Science and Technology, Vietnam Academy of Science and Technology, 18 Hoang Quoc Viet, Hanoi, Vietnam}
\author{Tran V. Huynh}
\affiliation{Graduate University of Science and Technology, Vietnam Academy of Science and Technology, 18 Hoang Quoc Viet, Hanoi, Vietnam}
\affiliation{University of Fire, 243 Khua Duy Tien, Hanoi, Vietnam}
\author{Vu D. Lam}
\affiliation{Graduate University of Science and Technology, Vietnam Academy of Science and Technology, 18 Hoang Quoc Viet, Hanoi, Vietnam}
\author{Anh-Tuan Le}
\affiliation{Phenikaa University Nano Institute (PHENA), Faculty of Materials Science and Engineering, Phenikaa University, Hanoi 12116, Vietnam}
\author{Katsunori Wakabayashi}
\affiliation{Department of Nanotechnology for Sustainable Energy, School of Science and Technology, Kwansei Gakuin University, Sanda, Hyogo 669-1337, Japan}
\date{\today}

\begin{abstract}
We theoretically investigate the plasmonic properties of mid-infrared graphene-based metamaterials and apply deep learning of a neural network for the inverse design. These artificial structures have square periodic arrays of graphene plasmonic resonators deposited on dielectric thin films. Optical spectra vary significantly with changes in structural parameters. To validate our theoretical approach, we carry out finite difference time domain simulations and compare computational results with theoretical calculations. Quantitatively good agreements among theoretical predictions, simulations and previous experiments allow us to employ this proposed theoretical model to generate reliable data for training and testing deep neural networks. By merging the pre-trained neural network with the inverse network, we implement calculations for inverse design of the graphene-based metameterials. We also discuss the limitation of the data-driven approach. 
\end{abstract}

\maketitle
\section{Introduction}
Metamaterials are artificial composites engineered to possess desired features not found in nature. Metals and dielectrics in metamaterials are periodically organized at the subwavelength scale. The incident light excites surface plasmons or collective oscillations of quasi-free electrons, which cause strong light-matter interactions. The subwavelength confinement of electromagnetic waves allow us to obtain negative refractive index \cite{9,10}, and perfect absorption and transmission \cite{11,12}. These properties have been applied to various areas including superlens \cite{13},  cloak of invisibility \cite{14}, sensing \cite{3,2}, and photothermal heating \cite{15,17,18,19}. However, plasmon lifetime in metal nanostructures is limited because of large inelastic losses of noble metals. The large Ohmic loss also reduces service life of optical confinement.

While graphene is a novel plasmonic material \cite{2,3,4}. Graphene can strongly confine electromagnetic fields, particularly in infrared regime, but dissipates a small amount of energy \cite{20}. A significant reduction of the heat dissipation in graphene compared to that in metals is caused by the small number of free electrons. One can easily tune optical and electrical properties of graphene via doping, applying external fields, and injecting charge carriers \cite{4}. Consequently, graphene-involved metamaterials are expected to possess various interesting behaviors. 

Graphene-based metamaterials can be investigated using different approaches. While experimental implementation and simulation are very expensive and time-consuming, theoretical approaches provide good insights into underlying mechanisms of metamaterials. Rapid collection of data from theoretical calculations, simulations, and experiments has introduced data-driven approaches to effectively investigate systems. Data-driven approaches using machine learning and deep learning have revolutionized plasmonic and photonic fields since they can speed up calculations and be one-time-cost approach after collecting data by expensive sources. Reliable theoretical models can generate huge amounts of systematic data for Machine Learning and Deep Learning analyses. Thus, combining theory and deep learning would pave the way for better understanding in science and introducing futuristic applications.

Inverse design problems have attracted scientific community since they allow us to accelerate the design process targeting desired properties \cite{31,32,6,21,33,34}. One has recently applied inverse design techniques to several physical systems such as quantum scattering theory \cite{31,21}, photonic devices \cite{32,6}, and thin film photovoltaic materials \cite{33}. Typically, inverse design problems are solved using the optimization approach in high-dimensional space, the genetic algorithm \cite{22}, and the adjoint method \cite{23}. Hower, these approaches require lengthy calculations and computational timescale. In this context, artificial neural networks provide faster calculations with higher precision.

In this work, we present a theoretical approach to investigate plasmonic properties of graphene-based nanostructures and use deep neural networks for the inverse design of the structure when knowing an optical spectrum. Our modeled systems mimic mid-infrared graphene detectors fabricated in Ref. \cite{2}. The validity of our theoretical approach is verified by comparisons with finite difference time domain (FDTD) simulations and the previous experimental work \cite{2}. Then, the theoretical calculations are used to generate training and testing data sets to train a neural network for forward prediction and inverse network. We also discuss limitations of these methods.

\section{Theoretical Background}
\subsection{Optical Response of Graphene-based Metamaterials}
\begin{figure}[htp]
\includegraphics[width=8.5cm]{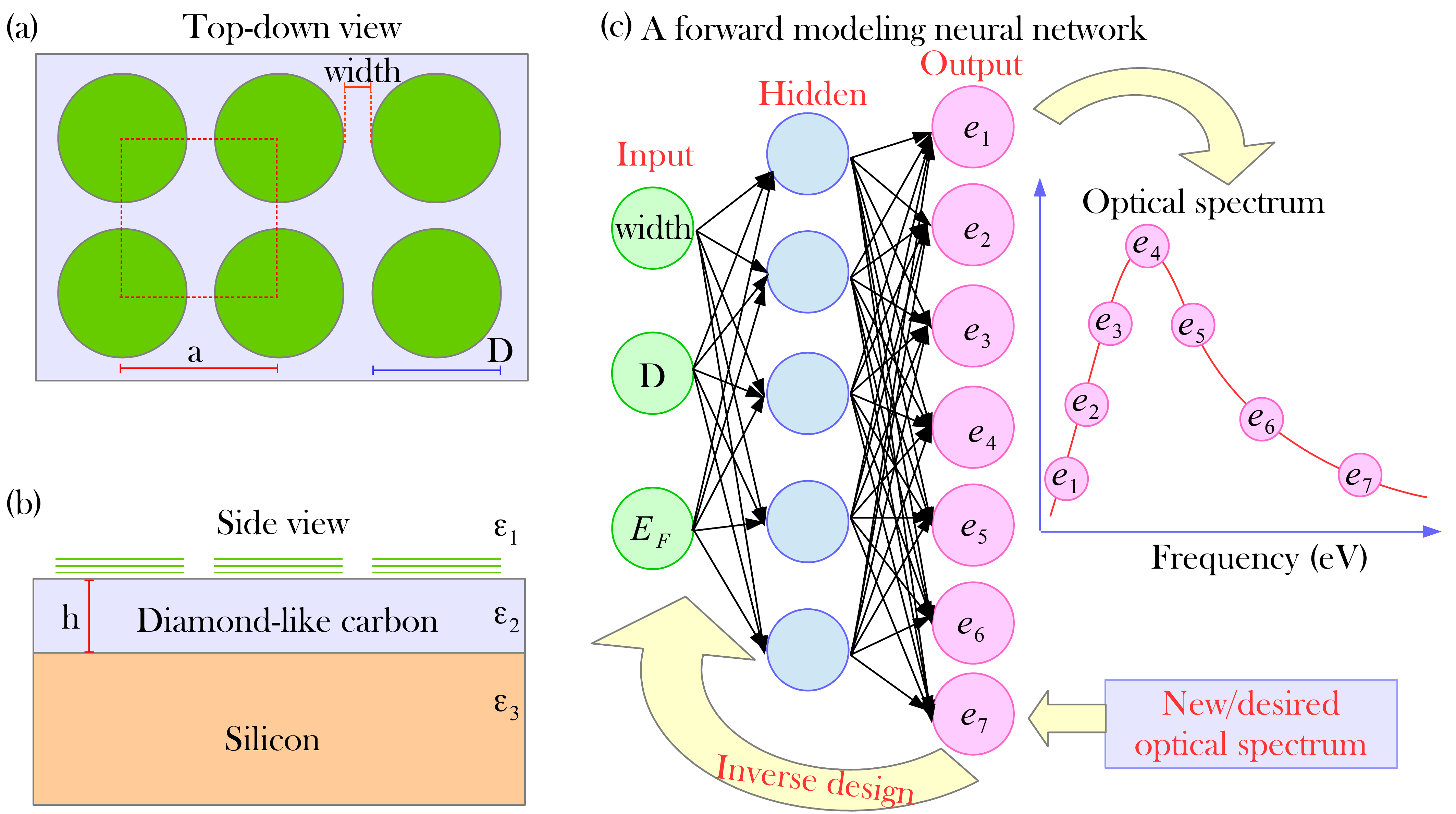}
\caption{\label{fig:1}(Color online) (a) The top-down view and (b) the side view of graphene based systems including structural parameters. (c) The neural network architecture has three main components: inputs (width, $D$, $E_F$), hidden layers, and outputs (the optical spectrum). \textcolor{red}{$E_F$ is a chemical potential of graphene.} There are three hidden layers in our actual neural network.}
\end{figure}

Figure \ref{fig:1}a and b show top-down and side view of our graphene-based systems in air medium ($\varepsilon_1=1$). The systems include a square lattice of graphene nanodisks on a diamond-like carbon thin film placed on a silicon substrate. The energy of optical phonon of the diamond-like carbon thin film is expected to be the same order as this energy of diamond (165 meV) \cite{40}. While the phonon energy of \ce{SiO_2} is approximately 55 meV \cite{39} and other conventional substrates have lower photon energy. Compared to other materials, the surface of diamond-like carbon films is non-polar and chemically inert due to very low trap density. The diamond-like carbon layer has a thickness of $h = 60$ nm and dielectric function $\varepsilon_2 = 6.25$. While the dielectric function of silicon is $\varepsilon_3 = 11.56$. The square lattice of graphene nanodisks has a lattice period, $a=270$ nm, a resonator size, $D=210$ nm, and a number of graphene resonator layers, $N$. The width between two adjacent graphene plasmons is $(a-D)$.

Under quasi-static approximations (the size is much smaller than incident wavelengths) associated with the dipole model, the polarizability of the graphene resonators as a function of frequency $\omega$ is analytically expressed as \cite{1,15,4,5}
\begin{eqnarray}
\alpha(\omega)=\frac{\varepsilon_1+\varepsilon_2}{2}D^3\frac{\zeta^2}{-i\omega D\cfrac{\varepsilon_1+\varepsilon_2}{2\sigma(\omega)}-\cfrac{1}{\eta}},
\label{eq:1}
\end{eqnarray}
where $\zeta$ and $\eta$ are geometric parameters and $\sigma(\omega)$ is the optical conducitivity of graphene plasmons. In Ref. \cite{1}, authors found $\zeta = 0.03801\exp\left(-8.569N d_g/D\right)-0.1108$ and $\eta=-0.01267\exp\left(-45.34N d_g/D\right)+0.8635$, here $d_g=0.334$ nm is the thickness of a graphene monolayer. The $N$-layers graphene conductivity in mid-infrared regime can be calculated using the random-phase approximation with zero-parallel wave vector \cite{4}. Typically, $\sigma(\omega)$ is contributed by both interband and intraband transitions. However, in the mid-infrared regime, the interband conductivity is ignored and we have
\begin{eqnarray}
\sigma(\omega)=\frac{Ne^2 i |E_F|}{\pi\hbar^2\left(\omega + i\tau^{-1} \right)},
\label{eq:2}
\end{eqnarray}
where $e$ is the electron charge, $\hbar$ is the reduced Planck constant, and $\tau$ is the carrier relaxation time. In our calculations, $\hbar\tau^{-1}=0.03$ eV. Note that Eq. (\ref{eq:2}) is the in-plane optical conductivity. The analytical expression only validates in the low frequency regime \cite{4}. Effective combinations of Eq. (\ref{eq:1}) and (\ref{eq:2}) require strong coupling condition \cite{41,42}, which is the ratio of the spacing distance between graphene disks to the diameter $D$ has to be very small. In our systems, graphene disk layers are separated by a dielectric layer \cite{42}. Thus, for simplicity, we assume the vertical optical conductivity is ignored. The stacking between graphene layers does not change the chemical potential and the horizontal optical conductivity is a simple addition of layers. 

The reflection and transmission coefficient of the graphene-based nanostructure are
\begin{eqnarray}
t_{13} &=& \frac{t_{12}t_{23}e^{i\left( \frac{\omega}{c}\sqrt{\varepsilon_2}h\right)}}{1+r_{12}r_{23}e^{2i\left( \frac{\omega}{c}\sqrt{\varepsilon_2}h\right)}},
\label{eq:3}
\end{eqnarray}
where
\begin{eqnarray}
r_0 &=& \frac{\sqrt{\varepsilon_2}-\sqrt{\varepsilon_1}}{\sqrt{\varepsilon_2}+\sqrt{\varepsilon_1}}, \quad t_0 = \frac{2\sqrt{\varepsilon_1}}{\sqrt{\varepsilon_2}+\sqrt{\varepsilon_1}}, \nonumber\\
r_{12} &=& r_0 - \frac{is(1-r_0)}{\alpha^{-1}-\gamma}, \quad t_{12} = t_0 + \frac{ist_0}{\alpha^{-1}-\gamma}, \nonumber\\
r_{23} &=& \frac{\sqrt{\varepsilon_3}-\sqrt{\varepsilon_{2}}}{\sqrt{\varepsilon_3}+\sqrt{\varepsilon_2}}, \quad t_{23} = \frac{2\sqrt{\varepsilon_2}}{\sqrt{\varepsilon_3}+\sqrt{\varepsilon_2}}, \nonumber\\
s &=& \frac{4\pi}{a^2}\frac{\omega/c}{\sqrt{\varepsilon_2}+\sqrt{\varepsilon_1}}, \quad \gamma \approx \frac{g}{a^3}\frac{2}{\varepsilon_1+\varepsilon_2}+is,
\label{eq:4}
\end{eqnarray}
where $r_{pq}$ and $t_{pq}$ are the bulk reflection and transmission coefficients, respectively, when electromagnetic fields strike from medium $p$ to $q$, $c$ is the speed of light, and $g\approx 4.52$ is the net dipolar interaction over the whole square lattice. From Eqs. (\ref{eq:3}) and (\ref{eq:4}), the transmission $|t_{13}|^2 \equiv |t_{13}(N)|^2$ for $N > 0$ and $N = 0$ corresponding to systems with and without graphene plasmonic resonators is calculated. In experiments, experimentalists measure the relative difference in these transmissions 1-$|t_{13}(N)|^2/t_{13}(N=0)|^2$ and call it the extinction spectrum. A variation of this spectrum determines graphene-plasmon-induced confinement of electromagnetic fields. 
\subsection{Simulation}
To validate our theoretical calculations, we use FDTD solver in Computer Simulation Technology (CST) microwave studio software \cite{35} to investigate the transmission properties and electromagnetically responses of the proposed metamaterial. The model of graphene in CST is described according to the Kubo formula. The dielectric function of multilayer graphene can be expressed as \cite{36,37,38}
\begin{eqnarray}
\varepsilon_{g}(\omega)= 1+\frac{i\sigma(\omega)}{\varepsilon_0\omega Nd_g}=1-\frac{e^2 |E_F|}{\pi\hbar^2\varepsilon_0\omega\left(\omega + i\tau^{-1} \right)d_g},
\label{eq:5}
\end{eqnarray}
where $\varepsilon_0$ is the vacuum permittivity. Equation (\ref{eq:5}) suggests an insensitivity of $\varepsilon_g(\omega)$ to the number of graphene layers. It means that the dielectric function of stacked multilayer graphene systems is approximately equal to that of monolayer graphene counterpart. This assumption is relatively reasonable since if we consider the stacked $N$-layer graphene disks as a film, its dielectric function is independent of the thickness.

In CST simulations, the incident electromagnetic wave is perpendicular to the surface, in which the electric and magnetic components are along $y$-axis and $x$-axis, respectively. We apply the open boundary condition along the $z$-axis while the periodic boundary conditions are employed along the $x$ and $y$ axes. Two transmitters and receivers are located on sides of the structure along the $z$-axis to measure transmission scattering parameters $S_{21}(\omega)$ of the electromagnetic wave when interacting with the graphene/DLC/silicon medium. Then, transmittance ($T$) would be obtained by $T(\omega)=\left[S_{21}(\omega)\right]^2$. Then, the extinction spectrum can be obtained by $1-T(N)/T_0$, where $T_0$ is the transmittance of the structure without graphene disks on top.

\subsection{Deep Neural Network}
Although the theoretical method has many advantages in understanding physical properties of graphene nanostructures, it is difficult to predict the structural parameters for a desired spectrum. We employ the tandem network introduced in Ref. \cite{6} for the inverse design of our graphene-based metamaterials having $N = 3$ in Fig. \ref{fig:1}a. 

\textcolor{red}{Inverse design is an ill-posed problem and many designs can be proposed to satisfy a given set of performance criteria. This is a research problem itself. There are two basic approaches. (1) The forward simulation/prediction is used to conjugate with several search techniques (for example, genetic algorithms, Bayesian optimization). If simulation is slow, this can be expensive and time-consuming. However, since machine learning always approximates accurate simulation, this can speed up calculations but reduce the accuracy. (2) Another approach is to use backward mapping from performance criteria back to design using machine learning models (mostly neural networks due to its flexibility). Finding multiple reasonable solutions is the most challenging problem if using this method. However, deep neural networks work very well when collecting a large amount of data, which are sequential and systematic. The calculations run very fast. In this work, the second method is employed.}

First, we use Eqs. (\ref{eq:3}) and (\ref{eq:4}) to generate roughly 11316 and 2860 extinction spectra for training and testing data sets, respectively. An entry of a theoretical calculation has three parameters. The diameter $D$ of a graphene nanodisk is constrained between 100 nm and 300 nm with a step size of 5 nm. While the width is changed from 10 nm to 120 nm with a step size of 5 nm. For the chemical potential $E_F$, we increase from 0.05 eV to 0.6 eV with a step size of 0.05 eV. \textcolor{red}{Practically, $E_F$ of graphene-based biosensors and photodetectors varies from 0.17 eV to 0.45 eV and the value can be controlled by an external electric field \cite{2,3,4}.} Each optical spectrum has 400 spectral points at a frequency range from 0.001 to 0.4 eV. 

Second, we use the training data set to train our neural network, which is depicted in Fig.\ref{fig:1}c. There are 6 hidden layers in the forward-modeling neural network. \textcolor{red}{Currently, no specific formula for choosing the number of hidden layers has been yet revealed. The running time grows with an increase of the number of hidden layers while the accuracy remains nearly unchanged as having more than three layers \cite{50}. In many studies, one empirically prefers six layers.} These layers sequentially have 1024-512-512-256-256-128 hidden units \textcolor{red}{\cite{6}. Again, there is also no universal role to estimate the number of units in each layer. In the computer-science point of view, the number of units is typically chosen as a power of two and the latter layer is equal or reduced by a factor of two in comparison with the previous one. Increasing the number of nodes gives more accurate but causes long-time calculations \cite{6}.} The learning rate is initialized at 0.0005. When finishing the training, a new set of structural parameters inputting to the forward network gives a predicted optical spectrum. 

Finally, a tandem neural network is formed by connecting the trained forward-modeling neural network to an inverse-design network with two hidden layers, which have 512 and 256 hidden units. The learning rate in the inverse design is set to 0.0001. This tandem neural network takes data points from a new/desired spectrum as an input to propose possible design parameters. Then, these design parameters are put in the trained forward-modeling neural network to generate the corresponding spectrum. The algorithm adjusts weights in the inverse network to minimize the mean square error between the real and predicted spectrum. The process is repeatedly carried out.

\section{Numerical Results and Discussions}
\begin{figure}[htp]
\includegraphics[width=8.4cm]{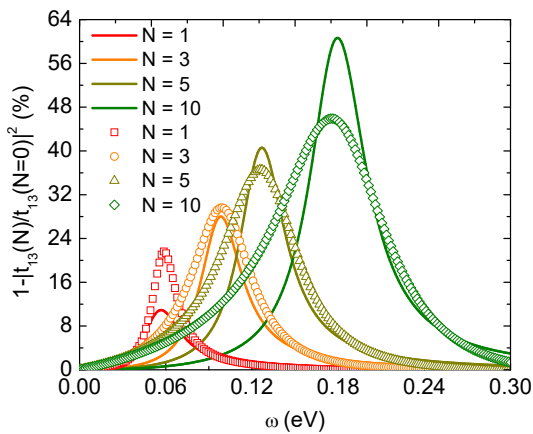}
\caption{\label{fig:4}(Color online) Extinction spectra for systems having a graphene-disk array with $D=210$ nm, the width of 60 nm, $E_F=0.45$ eV for the number of graphene layers $N = 1$, 3, 5 and 10 calculated using theoretical approach (solid lines) and CST simulations (points).}
\end{figure}

Figure \ref{fig:4} shows theoretical and simulation extinction spectra with several values of graphene plasmon layers ($N = 1$, 3, 5 and 10). Due to absorption of electromagnetic fields by surface plasmons in graphene nanodisks, the transmission is reduced. The reduction is enhanced when increasing the number of graphene layers and the plasmonic resonance is also blue-shifted. One can observe theoretical calculations quantitatively agree with simulations, particularly $N = 3$ and 5. The spectral positions of optical resonances predicted using these two approaches are very close. Particularly, the system with a square lattice of three-layers-graphene disks has the plasmonic peak at 0.1 eV ($\sim 806$ $\ce{cm^{-1}}$). This value is fully consistent with experiment in Ref. \cite{2}. The excellent agreement at $N = 3$ indicates that optical spectra calculated by our theoretical approach is reliable to generate for deep/machine learning study. Thus, in next calculations, we focus on the graphene-based metamaterials with $N = 3$.

\begin{figure}[htp]
\includegraphics[width=8.4cm]{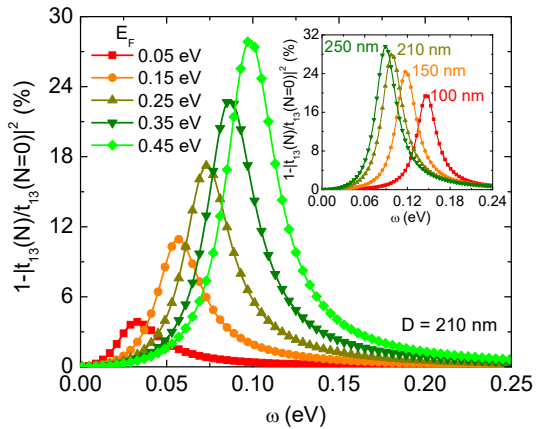}
\caption{\label{fig:2}(Color online) Theoretical extinction spectra for systems having a graphene-disk array with $D=210$ nm and the width of 60 nm at different values of the chemical potential $E_F$. The inset shows theoretical extinction spectra with $E_F=0.45$ eV and at several diameters of graphene nanodisks separated with their nearest neighbors by 60 nm.}
\end{figure}

The mainframe of Fig. \ref{fig:2} shows theoretical infrared extinction spectra of graphene-based systems with $D=210$ nm and the width of 60 nm at several values of the chemical potentials. The calculations are carried out using Eqs. (\ref{eq:3}) and (\ref{eq:4}). For $E_F=0.45$ eV, all structural parameters are identical to the fabricated detector in Ref. \cite{2}. The plasmonic peak is roughly located at 0.1 eV. The value is in a quantitative accordance with the prior experimental result \cite{2}. While a decrease of $E_F$ not only red-shifts the surface plasmon resonance, but also significantly reduces an amplitude of the optical signal. The reason is $\sigma(\omega) \sim E_F$. The metal-like or plasmonic properties are lost with decreasing the chemical potential through reducing the numnber of free electrons.

The inset of Fig. \ref{fig:2} shows the sensitivity of optical spectra to the diameter of graphene nanodisks. The nearest distance between two plasmonic resonators is fixed at 60 nm. Since $\alpha(\omega) \sim D^3$, the extinction cross section of a graphene nanodisk approximated by $4 \ce{Im}(\alpha)\omega/c$ is proportional to $D^3$. Thus, a decrease of the diameter weakens the plasmonic coupling among resonators and lowers the optical peak. In addition, one can observe the blue-shifts in the spectrum when reducing the size of nanodisks. These behaviors suggest increasing the size of graphene plasmons enables to confine more mid-infrared optical energy. The trapped energy is highly localized in plasmonic resonators and thermally dissipates through the system.

\begin{figure}[htp]
\includegraphics[width=8.4cm]{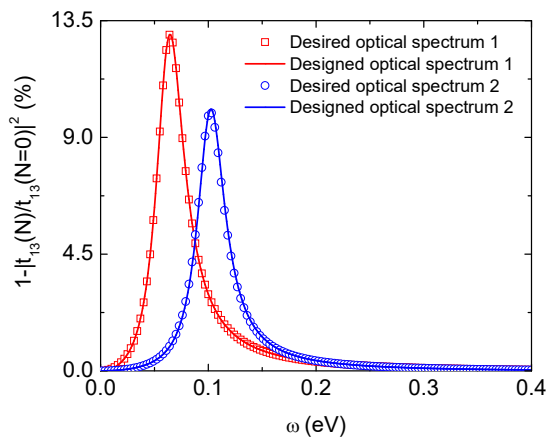}
\caption{\label{fig:3}(Color online) Two examples of the inverse design using the tandem neural network. The solid curves and data points correspond to the predicted spectrum and target/desired spectrum, respectively.}
\end{figure}

Once the neural network is trained by our training and testing data sets, it can very fast calculate a new extinction spectrum when inputting a new set of structural parameters including $D$, the width, and $E_F$. The trained network is now incorporated with the inverse network to construct the tandem network for improving accuracy of inverse design. To demonstrate the validity of our deep neural network, we \textcolor{red}{exame this network by predicting structural parameters for two desired spectra and show results in Fig. \ref{fig:3}. First, equations (\ref{eq:3}) and (\ref{eq:4}) are used to generate two target optical spectra 1 and 2 with structural parameters (247 nm, 103 nm, 0.22 eV) and (127 nm, 103 nm, 0.22 eV), respectively. The plasmonic peaks are approximately located at $0.064$ eV $\approx 19372.53$ nm and $0.102$ eV $\approx 12155.31$ nm. From this, one finds that difference in the position of the plasmon peak is 0.038 eV or 7217.22 nm. Second, these two spectra are learned by our neural network and the calculations give parameter sets ($D$, width, $E_F$) = (206 nm, 64 nm, 0.188 eV) and (130 nm, 108 nm, 0.279 eV) corresponding to these spectrum 1 and 2, respectively. Then, the optical spectra are calculated using our Eqs.(\ref{eq:3}) and (\ref{eq:4}) associated with ($D$, width, $E_F$) = (206 nm, 64 nm, 0.188 eV) and (130 nm, 108 nm, 0.279 eV). As shown in Fig.\ref{fig:3}, perfect overlaps between these calculated spectra and the target counterparts are found. These findings reveal the spectrum 1 can be obtained from two parameter sets ($D$, width, $E_F$) = (206 nm, 64 nm, 0.188 eV) and (247 nm, 103 nm, 0.22 eV). While graphene-based metamaterials having ($D$, width, $E_F$) = (130 nm, 108 nm, 0.279 eV) and (127 nm, 103 nm, 0.22 eV) can provide the spectrum 2. These quantitative good agreements} clearly validate our designs and the data-driven approach for inverse design.

\textcolor{red}{Again,} one of the most challenging problems in the inverse design is nonuniqueness. There may be many designs having an identical performance. It is difficult to overcome this issue. In a recent work \cite{6}, Liu and his coworkers proposed the tandem architecture to handle the nonunique designs. However, our \textcolor{red}{above} calculations numerically prove that it is not universal. Clearly, the real and deep-learning-predicted structures are different. Although this result does not negate the validity of the tandem neural network for the inverse design problem, it indicates that the nonunique issue is still not handled. \textcolor{red}{Furthermore, since the structure-property relationship is highly non-linear, calculations using this tandem neural network can work well with high dimensional data features as stated in Ref. \cite{6}, but it does not implies the success happens at the low dimensional one.}

\section{Conclusions}
We have investigated extinction spectra of graphene-based metamaterials using theory, simulations, and deep learning. The nanostructures have a square array of three-layers graphene nanodisks placed on the diamond-like carbon thin film on a semi-infinite silicon substrate. The polarizability of array of graphene nanodisks is calculated using the dipole model associated with the random-phase approximation. Based on this analysis, we analytically express the transmission coefficient and calculate the extinction spectra as a function of many structural parameters. The numerical results agree quantitatively well with CST simulations. Since the FDTD simulations require very large computational work and running time, the theoretical calculations are employed to generate reliable data for the tandem neural network. After training the artificial neural network, it has been used to solve the inverse design problem. Our deep learning calculations have showed that the predicted design can be accurately given for a target performance. However, calculations based on the tandem neural network do not handle the issue of nonunique design.  
\begin{acknowledgments}
This work was supported by JSPS KAKENHI Grant Numbers JP19F18322 and JP18H01154. This research is funded by the joint research project between the Vietnam National Foundation for Science and Technology Development (NAFOSTED) and the Research Foundation Flanders (FWO) under grant number FWO.103.2017.01.
\end{acknowledgments}

\end{document}